\documentclass[aps,pre,twocolumn,groupedaddress,floatfix,longbibliography,superscriptaddress]{revtex4-1}

\usepackage{amsmath}
\usepackage{amssymb}
\usepackage{amsfonts}
\usepackage{graphicx}
\usepackage{bbold}
\usepackage{bm}
\usepackage{bm}
\usepackage{cancel}
\usepackage{times,float}
\usepackage{graphicx}
\usepackage[usenames,dvipsnames,svgnames]{xcolor}
\usepackage{hyperref}
\hypersetup{colorlinks=true, linkcolor=Blue, citecolor=Blue,urlcolor=Blue}
\usepackage{multirow}
\usepackage{ulem}
\usepackage{color,epsfig}
\usepackage{bm}
\usepackage{slashed}
\usepackage{hyperref}
\usepackage{subfigure}
\usepackage{feynmf}
\usepackage{mathrsfs}
\usepackage{simpler-wick}

\newcommand{\bea}{\begin{eqnarray}}
\newcommand{\eea}{\end{eqnarray}}
\newcommand{\be}{\begin{equation}}
\newcommand{\ee}{\end{equation}}

\newcommand{\nn}{\nonumber}
\newcommand{\ii}{\mathrm{i}}

\begin{document}

\title{Temperature fluctuations in a relativistic gas: Pressure corrections and possible consequences in the deconfinement transition}
\author{Jorge David Casta\~no-Yepes}
\email{jcastano@uc.cl}
\affiliation{Facultad de F\'isica, Pontificia Universidad Cat\'olica de Chile, Vicu\~{n}a Mackenna 4860, Santiago, Chile}
\author{Marcelo Loewe}
\email{mloewe@fis.puc.cl}
%\affiliation{Facultad de F\'isica, Pontificia Universidad Cat\'olica de Chile, Vicu\~{n}a Mackenna 4860, Santiago, Chile}
\affiliation{Centre for Theoretical and Mathematical Physics, and Department of Physics, University of Cape Town, Rondebosch 7700, South Africa}
\affiliation{Facultad de Ingeniería, Arquitectura y Diseño, Universidad San Sebastián, Santiago, Chile}
\author{Enrique Mu\~noz}
\email{ejmunoz@uc.cl}
\affiliation{Facultad de F\'isica, Pontificia Universidad Cat\'olica de Chile, Vicu\~{n}a Mackenna 4860, Santiago, Chile}
\affiliation{Center for Nanotechnology and Advanced Materials CIEN-UC, Avenida Vicuña Mackenna 4860, Santiago, Chile}
\author{Juan Crist\'obal Rojas}
\email{jurojas@ucn.cl}
\affiliation{Departamento de Física, Universidad Cat\'olica del Norte, Angamos 610, Antofagasta, Chile}
%\date{ }

\begin{abstract}
In this work, we study the effects of random temperature fluctuations on the equation of state of a non-interacting, relativistic fermion gas by means of the {\it replica method}. This picture provides a conceptual model for a non-equilibrium system, depicted as an ensemble of subsystems at different temperatures, randomly distributed with respect to a given mean value.
We then assume the temperature displays stochastic fluctuations $T = T_0 + \delta T$ with respect to its ensemble average value $T_0$, with zero mean $\overline{\delta T} = 0$ and standard deviation $\overline{\delta T^2} = \Delta$. By means of the replica method, we obtain the average grand canonical potential, leading to the equation of state of the fermion gas expressed in terms of the excess pressure caused by these fluctuations with respect to the ideal gas at uniform temperature. We further extend our results for the ideal Bose gas as well. Our findings reveal an increase in pressure as the system's ensemble average temperature $T_0$ rises, consistently exceeding the pressure observed in an equilibrium state. Finally, we explore the implications for the deconfinement transition in the context of the simple Bag model, where we show that the critical temperature decreases.

\end{abstract}

\maketitle 

%----------------------------------------------------------------
\section{Introduction}
%----------------------------------------------------------------
Finite temperature quantum field theory provides a general framework to study the statistical and thermodynamic properties of quantum matter, from condensed matter systems to applications in high-energy physics~\cite{altland2010condensed,le2000thermal,kapusta2007finite}. However, many experimental systems of interest are not in strict thermodynamic equilibrium. For instance, in high-energy experiments such as ultra-relativistic heavy-ion collisions, the emergence of the quark-gluon plasma is a broadly accepted phenomenon~\cite{ARSENE20051,ADCOX2005184,Becattini_2014}. This state involves the coexistence of the fundamental degrees of freedom in Quantum Chromodynamics, in principle assumed to be in thermal equilibrium, so that finite temperature field theory has been instrumental in successfully explaining and predicting a wide array of observables arising from such systems. Nevertheless, the initial and hadronization phases of a heavy-ion collision are not in thermal equilibrium~\cite{PhysRevC.83.034907,BLAIZOT1987847,PhysRevC.83.044903,RafelskyHadronization}. This leads to essential questions: How do thermal fluctuations impact observables during these stages? And what theoretical approaches can correctly capture those effects?

A number of theoretical approximations have been developed to represent non-equilibrium conditions in quantum systems. For instance, the Keldysh contour path emerged as a formalism to describe the quantum mechanical evolution of systems under time-varying external fields~\cite{doi:10.1142/9789811279461_0007}. This formalism has found notable applications, particularly within strongly correlated electron systems~\cite{e20050366,PhysRevLett.110.016601,Muñoz_2017,PhysRevB.98.195430,MuñozBook}, offering insights into their many-body properties and non-equilibrium dynamics~\cite{RevModPhys.86.779,Sieberer_2016}. Another approach used to model the fluctuations of intensive thermodynamic parameters is the so-called \textit{superstatistics}. This method assumes an ensemble of subsystems, each of them individually in local thermal equilibrium~\cite{beck2003superstatistics}. Although this approximation represents a semi-classical description of the thermal fluctuations, it has been applied extensively in condensed matter systems~\cite{doi:10.1142/S0217732319500238,castano2020super}, information theory~\cite{PhysRevE.95.042111,PhysRevE.104.024139,Castaño-Yepes2022}, and more recently to the study of the QCD phase diagram~\cite{ayala2018superstatistics,PhysRevD.106.116019}.

There is an alternative perspective to address fluctuations or disorder within a system: the well-known {\it replica trick}, introduced by Parisi as a method to average the free energy, defined via the logarithm of the partition function $\ln Z$, of a system over quenched (or frozen) disorder~\cite{doi:10.1142/0271}. This method builds on the mathematical identity
\begin{equation}
\overline{\ln Z} = \lim_{n\to0}\frac{\overline{Z^n}-1}{n},
\label{eq_replica_trick}
\end{equation}
which in practical implementation involves replicating the system $n$-times, with the corresponding partition function as an effective coupling of the $n$-replicas of the same Lagrangian. This procedure can be performed either at the level of the Canonical Ensemble, leading to an statistical average of the Helmholtz free-energy $\overline{\mathcal{F}}(\mathcal{N},\mathcal{V},T) = -  T\, \overline{\ln Z}(\mathcal{N},\mathcal{V},T)$ or the Grand Canonical Ensemble, where in such case one obtains the corresponding average of the Grand Potential $\overline{\Omega}(\mu,\mathcal{V},T) = -  T\, \overline{\ln Z}(\mu,\mathcal{V},T)$.

In a series of two recent articles~\cite{PhysRevD.107.096014,PhysRevD.108.116013}, we applied this formalism to investigate the effects of stochastic fluctuations in a classical background magnetic field on the properties of a Quantum Electrodynamics (QED) medium. Specifically, we showed that magnetic fluctuations lead to an effective interaction between the fermions. This approach, at the perturbative level, results in quasi-particles propagating through a dispersive medium~\cite{PhysRevD.107.096014}. Moreover, our mean-field analysis predicted the emergence of order parameters representing the components of a vector current~\cite{PhysRevD.108.116013}. Both perspectives revealed deviations from $U(1)$ symmetry, highlighting the influence of stochastic fluctuations within the background magnetic field on the properties of the QED medium.

In this study, we shall assume a non-equilibrium scenario, where temperature is then not defined uniformly through the whole system, but smaller regions may still be pictured as nearly-thermalized subsystems. Therefore, we model this situation by an ensemble of subsystems whose individual temperatures $T = T_0 + \delta T$ are subjected to stochastic fluctuations with zero mean $\overline{\delta T} = 0$, but finite variance $\overline{\delta T^2} = \Delta$.

For technical reasons, it is more convenient to represent these fluctuations in terms of the inverse temperature 
\bea
\beta &=& (T_0 + \delta T)^{-1}= T_0^{-1} - \frac{\delta T}{T_0^2} = \beta_0 + \delta\beta,
\eea
where clearly we have the corresponding relations
\bea
\delta\beta &=& - \frac{\delta T}{T_0^2}\nn\\
\overline{\delta\beta} &=& -T_0^{-2}\overline{\delta T} = 0\nn\\
\overline{\delta\beta^2} &=& T_0^{-4}\overline{\delta T^2} = \beta_0^4\Delta = \Delta_{\beta}.
\eea

We capture these statistical features by assuming a Gaussian distribution with zero mean, i.e.
\be
dP[\delta\beta] = \frac{d(\delta\beta)}{\sqrt{2\pi\Delta_{\beta}}} e^{-\frac{\delta\beta^2}{2\Delta_{\beta}}}.
\label{eq_dP}
\ee

Therefore, the corresponding moments of the thermal fluctuations are given by the exact expressions ( $\forall\,\,j \in \mathbb{N}$)
\bea
\overline{\delta\beta} &=& 
\overline{\delta\beta^{2 j-1}} = 0,\nn\\
\overline{\delta\beta^{2}} &=& \Delta_{\beta},\nn\\
\overline{\delta\beta^{2 j}} &=& \Delta_{\beta}^{j}(2j-1)!!.
\label{eq_beta_stat}
\eea

We will further apply the replica formalism to average over these temperature fluctuations. Subsequently, in the Grand-Canonical ensemble at finite chemical potential, by using the Matsubara imaginary time formalism, we calculate the system's Grand Potential, and its equation of state (EoS).

%----------------------------------------------------------------
\section{The Grand Canonical partition function}
%----------------------------------------------------------------

Let us start by considering the definition of the partition function in the Grand-Canonical ensemble
\be
Z(\mu,\mathcal{V},T) = {\rm{Tr}} \left[ e^{-\beta \left( \hat{H} -\mu \hat{N} \right)} \right]
\ee
where, in particular, we shall focus on a system of QED fermions, described by the Hamiltonian operator (including the chemical potential)
\bea
\hat{H} -\mu \hat{N} &=& \int d^{3}x\,\hat{\psi}^{\dagger}(\mathbf{x})\gamma^0\left[ \boldsymbol{\gamma}\cdot(-\ii \nabla) + m - \gamma^0\mu\right]\hat{\psi}(\mathbf{x})\nn\\
&\equiv& \hat{K}.
\label{eq_K}
\eea

As we stated in the Introduction, we shall assume that our system is not fully thermalized, but for a quenched distribution of local temperatures, with the statistical properties described by Eq.~\eqref{eq_beta_stat}. Moreover, the statistical average over such distribution of temperatures is calculated via the replica trick~\ref{eq_replica_trick}
\bea
\overline{\ln Z} = \lim_{n\rightarrow 0}\frac{1}{n} \left( \overline{{\rm{Tr}}\left[ \exp\left\{-(\beta_0 + \delta\beta)\sum_{a=1}^{n}\hat{K}^{(a)}\right\}\right]} - 1 \right),
\label{eq_LNZ1}
\eea
where here each replica $1\le a \le n$ has an associated operator $\hat{K}^{(a)}$ as defined in Eq.~\eqref{eq_K}, for fermion field operators with
and additional index $\hat{\psi}_a (\mathbf{x})$.

Let us now expand the exponential inside the trace in Eq.~\eqref{eq_LNZ1} in powers of the fluctuation $\delta\beta$, and then take the statistical average of each term using the properties of the distribution in Eq.~\eqref{eq_beta_stat},
as follows
\begin{widetext}
\bea
\overline {Z^{n}} = \overline{{\rm{Tr}}\left[ \exp\left\{-(\beta_0 + \delta\beta)\sum_{a=1}^{n}\hat{K}^{(a)}\right\}\right]}
&=& \int dP[\delta\beta] {\rm{Tr}}\left[ e^{-\beta_0  \sum_{a=1}^{n}\hat{K}^{(a)}}\left(1 +  \sum_{j=1}^{\infty}\frac{(-1)^j\left(\delta\beta\right)^j}{j!} \left(\sum_{a=1}^{n}\hat{K}^{(a)} \right)^j\right)\right]\nn\\
&=& {\rm{Tr}}\left[ e^{-\beta_0  \sum_{a=1}^{n}\hat{K}^{(a)}}\left(1 +\sum_{j=1}^{\infty}\frac{\Delta_{\beta}^j}{(2j)!}(2j-1)!!\left(\sum_{a=1}^{n}\hat{K}^{(a)}\right)^{2j}\right)\right]\nn\\
&=& \left(1 + \sum_{j=1}^{\infty}\frac{\Delta_{\beta}^j}{(2j)!}(2j-1)!!\frac{\partial^{2j}}{\partial\beta_0^{2j}}\right) {\rm{Tr}}\left[ e^{-\beta_0  \sum_{a=1}^{n}\hat{K}^{(a)}}\right].
\eea
\end{widetext}

Remarkably, after the statistical average over fluctuations was taken, the power expansion involving the trace can be expressed as temperature derivatives of the partition function of the reference system, as follows
\bea
\overline{Z^n} = \left(1 + \sum_{j=1}^{\infty}\frac{\Delta_{\beta}^j}{(2j)!}(2j-1)!!\frac{\partial^{2j}}{\partial\beta_0^{2j}}\right) Z_0^n
\label{eq_Zn}
\eea
where we defined the grand canonical partition function for the free reference system as
\be
Z_0^n = {\rm{Tr}}\left[ e^{-\beta_0  \sum_{a=1}^{n}\hat{K}^{(a)}}\right].
\ee

This is nothing but n-replicas of an ideal gas of relativistic fermions, and hence its partition function is calculated from the standard functional integral representation
\begin{widetext}
\bea
Z_0^{n} &=& \prod_{a=1}^{n}\int\mathcal{D}\left[ \psi_a^{\dagger},\psi_a \right]
\exp\left[-\int_0^{\beta_0}d\tau
\sum_{a=1}^n \psi_a^{\dagger}(\mathbf{x},\tau)\gamma^{0}\left(
\gamma^0\left(\partial_{\tau} - \mu\right) + \mathbf{\gamma}\cdot\mathbf{p} + m
\right)\psi_a(\mathbf{x},\tau)
\right]\nn\\
&=& {\rm{det}}\left[\partial_{\tau} - \mu + \gamma^0\mathbf{\gamma}\cdot\mathbf{p} + m\gamma^0
 \right]^n\nn\\
&=& \exp\left\{n{\rm{Tr}}\ln\left[ 
\partial_{\tau} - \mu + \gamma^0\mathbf{\gamma}\cdot\mathbf{p} + m\gamma^0
\right] \right\}\nn\\
&=& \exp{\left(n\ln Z_0\right)},
\label{eq_LnZ0_Funcint}
\eea
\end{widetext}
where the symbol ${\rm{Tr}}$ stands for the functional trace (integral over phase-space) and the trace in the space of Dirac matrices. Here, we also defined the partition function for the fermion gas
\bea
\ln Z_0 &=& {\rm{Tr}}\ln\left[ 
\partial_{\tau} - \mu + \gamma^0\mathbf{\gamma}\cdot\mathbf{p} + m\gamma^0
\right]\\
&=& \mathcal{V}\int\frac{d^3 p}{(2\pi)^3}\sum_{k\in\mathbb{Z}}{\rm{tr}}\ln\left[ \ii\omega_k - \mu + \gamma^0\mathbf{\gamma}\cdot\mathbf{p} + m\gamma^0\right],\nn
\label{eq_LnZ0}
\eea
where we diagonalized the operator in Matsubara-momentum space, and $\omega_k = (2 k +1)\pi/\beta_0$ ($k\in\mathbb{Z}$) are the Fermi Matsubara frequencies. Using the elementary identity, valid for any diagonalizable matrix $\hat{A}$,
\bea
{\rm{tr}}\ln\hat{A} &=& {\rm{tr}}\left[\hat{P}^{-1}(\ln \hat{A}) \hat{P}\right] = {\rm{tr}}\left[\ln (\hat{P}^{-1}\hat{A}\hat{P})\right]\nn\\ 
&=&  \sum_{i}\ln\lambda_i,
\eea
where $\hat{P}$ is the unitary transformation that diagonalizes $\hat{A}$ and $\lambda_i$ its eigenvalues.

We evaluate the trace in Eq.~\eqref{eq_LnZ0} from the (double-degenerate) eigenvalues of the matrix in the argument of the logarithm: $\ii\omega_k - \mu \pm E_\mathbf{p}$, with $E_\mathbf{p} = \sqrt{p^2 + m^2}$, to obtain
\bea
\ln Z_0 
&=& 2\mathcal{V}\int\frac{d^3 p}{(2\pi)^3}\sum_{k\in\mathbb{Z}}\left\{\ln\left[ \ii\omega_k - \mu + E_\mathbf{p}\right] \right.\nn\\
&&\left.+ \ln\left[ \ii\omega_k - \mu - E_\mathbf{p}\right]\right\}\nn\\
&=& 2\mathcal{V}\int\frac{d^3 p}{(2\pi)^3}\left\{
\ln\left( 1 + e^{\beta_0(\mu-E_\mathbf{p} )} \right)\right.\nn\\ 
&&\left.+ \ln\left( 1 + e^{\beta_0(\mu+E_\mathbf{p} )} \right)
\right\}.
\eea

The evaluation of the Matsubara sum in the final step is presented in detail in Appendix~\ref{app_Matsubara}.
Finally, inserting this result into Eq.~\eqref{eq_LnZ0_Funcint}, and into Eq.~\eqref{eq_LNZ1}, we obtain
\begin{widetext}
\bea
\overline{\ln Z} &=& \lim_{n\rightarrow 0}\frac{\overline{Z^{n}}-1}{n}= \left(1 + \sum_{j=1}^{\infty}\frac{\Delta_{\beta}^j}{(2j)!}(2j-1)!!\frac{\partial^{2j}}{\partial\beta_0^{2j}}\right) \lim_{n\rightarrow 0}\frac{e^{n\ln Z_0}-1}{n}\nn\\
&=& \left(1 + \sum_{j=1}^{\infty}\frac{(\Delta_{\beta}/2)^j}{j!}\frac{\partial^{2j}}{\partial\beta_0^{2j}}\right) \ln Z_0\nn\\
&=& \exp\left[\frac{\Delta_{\beta}}{2}\frac{\partial^2}{\partial\beta_0^2} \right]\ln Z_0,
\label{eq_EOS}
\eea
\end{widetext}
where in the second line we used the identity $(2j-1)!!/(2j)! = 2^{-j}/j!$, and we finally reassembled the series in the form of the exponential differential operator.
%----------------------------------------------------------------
\section{The Equation of state}
%----------------------------------------------------------------
\begin{figure*}
    \centering
    \includegraphics[scale=0.8]{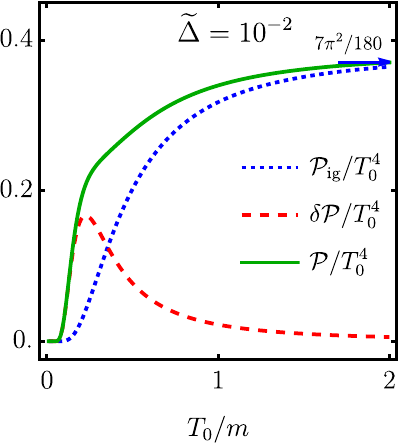}\hspace{1.3cm}
    \includegraphics[scale=0.8]{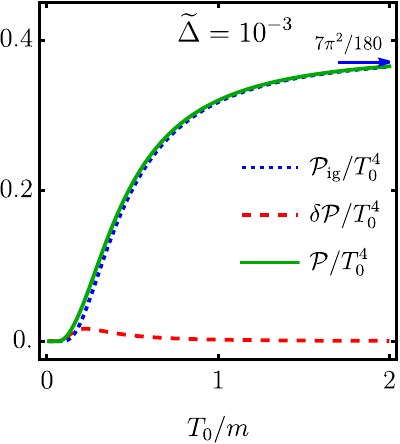}
    \caption{Pressure normalized to $T_0^4$ when $\mu=0$ for the ideal fermion gas (dotted line), the excess pressure of Eq.~\eqref{eq:dP} (dashed line), and the total pressure (continuous line). The arrow indicates the asymptotic ideal gas limit at high temperatures. }
    \label{fig:dPa_dPb}
\end{figure*}

%\begin{figure}
%    \centering
%   \includegraphics[scale=0.7]{dP1}
%    \caption{{\bf{\color{red} propuesta 1:}}Excess pressure computed up to order $O(\Delta)$, as a function of the temperature $T_0$, and the chemical potential $\mu$.}
%    \label{fig:dP}
%\end{figure}

\begin{figure}
    \centering
    \includegraphics[scale=0.7]{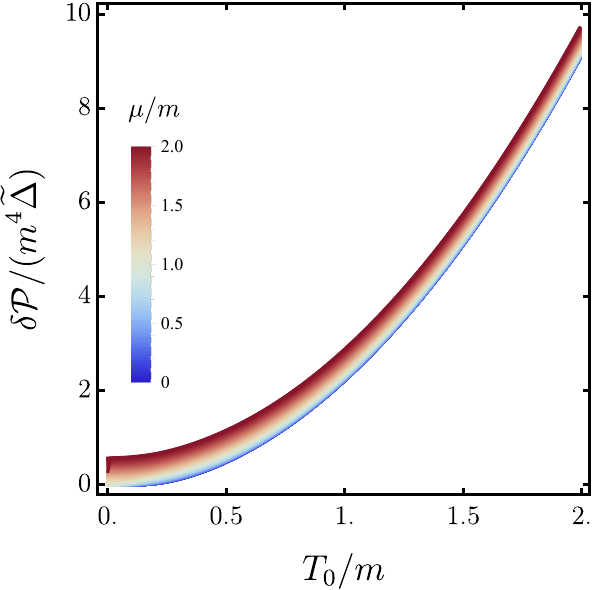}
    \caption{Excess pressure of the Fermi gas, computed up to order $O(\Delta)$ from Eq.~\eqref{eq:dP}, as a function of the average temperature $T_0$, and the chemical potential $\mu$.}
    \label{fig:dP2}
\end{figure}

Even though Eq.~\eqref{eq_EOS} is an exact result, for the purposes of this work, we shall assume that the temperature fluctuations are weak, in the sense $m^2\Delta_{\beta} \ll 1$. Therefore, we shall truncate the expansion up to first order in the fluctuation $\Delta$, to obtain
\bea
\overline{\ln Z/Z_0} &=& \frac{\Delta_{\beta}}{2}\frac{\partial^2}{\partial\beta_0^2}\ln Z_0 + O(\Delta_{\beta}^2)\nn\\
&=& \beta_0\left(\mathcal{P}\mathcal{V} - \left( \mathcal{P}\mathcal{V}\right)_{ig}\right),
\eea
where $\beta_0 \left(\mathcal{P} \mathcal{V}\right)_{ig} = \ln Z_0$ is the equation of state for the ideal Fermi gas. Therefore, up to $O(\Delta^2)$, the {\it{excess pressure}} $\delta\mathcal{P}\equiv \mathcal{P} - \mathcal{P}_\text{ig}$ of the Fermi gas due to the average effect of the temperature fluctuation is
\bea
\delta \mathcal{P}&\equiv& \mathcal{P} - \mathcal{P}_\text{ig}= \frac{\Delta_{\beta}}{2\mathcal{V}\beta_0}\frac{\partial^2}{\partial\beta_0^2}\ln Z_0 + O(\Delta^2).
\label{eq_dPf}
\eea

Before calculating explicitly this contribution, it is interesting to analyze its physical interpretation using general Thermodynamic relations. From the differential form of the Grand Potential for the ideal reference system, we have
\bea
d\Omega_{0} = -\mathcal{P}d\mathcal{V} - S dT_0 - N d\mu,
\eea
from which we conclude that
\be
S = -\left.\frac{\partial\Omega_0}{\partial T_0}\right|_{\mu,\mathcal{V}}.
\ee

On the other hand, using $T_0 = \beta_0^{-1}$ and the definition $\Omega_0 = -T_0\ln Z_0$, it is possible to show the identity (see Appendix~\ref{app_thermo} for details)
\bea
\frac{\Delta_{\beta}}{2\beta_0\mathcal{V}}\frac{\partial^2}{\partial\beta_0^2}\ln Z_0 &=& \frac{\Delta_{\beta} T_0^2}{2\mathcal{V}}T_0\left.\frac{\partial S}{\partial T_0}\right|_{\mu,\mathcal{V}}\nn\\
&=& \frac{\Delta_{\beta} T_0^2}{2\mathcal{V}}\left[ C_{v} + \frac{\left( T_0\,\left.\frac{\partial N}{\partial T_0 }\right|_{\mu,\mathcal{V}}\right)^2}{\langle (\delta\hat{N})^2\rangle} \right]\nn\\
&\ge& 0.
\label{eq_ineq}
\eea

Therefore, based on general thermodynamics considerations for the ideal reference system, we expect for the excess pressure  due to random temperature fluctuations in the ensemble to be positive $\delta\mathcal{P}\ge 0$ at first order in $\Delta$. Indeed, a direct calculation of the second derivative of the Grand Partition function, leading to the explicit formula
\bea
\delta \mathcal{P}&=& \frac{\Delta_{\beta}}{\beta_0}\sum_{s=\pm1}\int \frac{d^3 p}{(2\pi)^3} 
(E_\mathbf{p}+ s\mu)^2 n_\text{F}\left(\frac{E_\mathbf{p} +s\mu}{T_0}\right)\nn\\
&&\quad\quad\quad\quad\quad\quad\quad\times\left[1 - n_\text{F}\left(\frac{E_\mathbf{p} +s\mu}{T_0}\right)\right],
\eea
where $n_F(x) = \left( e^x + 1\right)^{-1}$ is the Fermi distribution, allows us to verify that it is clearly positive definite.

In order to evaluate the integral, it is convenient to perfom the change of variable (here $E=E_\mathbf{p}$ for short notation)
\bea
\mathbf{p}^2=E^2-m^2
%&\to& dp=\frac{EdE}{\sqrt{E^2-m^2}}\nn\\
\to d^3p=4\pi p^2dp=4\pi\sqrt{E^2-m^2}EdE,\nn\\
\eea
and later define the dimensionless variables $x\equiv E/m$, $y\equiv T_0/m$, $z=\mu/m$, and $\widetilde{\Delta}\equiv\Delta/m^2$, such that we have
\bea
\delta P&=&\frac{m^4\widetilde{\Delta} }{2\pi^2 y^3}\sum_{s=\pm1}\int_1^\infty dx~x(x+sz)^2\sqrt{x^2-1}\nn\\
&&\quad\quad\quad\quad\times n_\text{F}\left(\frac{x+sz}{y}\right)n_\text{F}\left(-\frac{x+sz}{y}\right),
\label{eq:dP}
\eea
where we used the property $n_\text{F}(-x)=1-n_\text{F}(x)$.

We represent the ratio between the total ensemble-average pressure $\mathcal{P} = \mathcal{P}_\text{ig} + \delta\mathcal{P}$ and the fourth power of the average temperature, i.e. $\mathcal{P}/T_0^4$ in Figure~\ref{fig:dPa_dPb}, for the specific case of vanishing chemical potential. For the sake of comparison, we included the well known temperature dependence of the ideal relativistic Fermi gas, that asymptotically attains the limit $\mathcal{P}/T_0^4 \sim 7\pi^2/180$ as $T_0/m > 1$~\cite{le2000thermal}. We also represent the excess pressure contribution $\delta\mathcal{P}$ arising from stochastic fluctuations in the temperature, calculated after Eq.~\eqref{eq:dP}. As can be noticed, the excess pressure represents a significant contribution at low equilibrium temperatures $T_0\lesssim m$, but it rapidly decreases at higher temperatures $T_0 > m$. Moreover, the magnitude of this deviation from the ideal gas pressure is proportional to the parameter $\widetilde{\Delta}$ representing the standard deviation in the temperature distribution accross the ensemble of subsystems. 

Following the same considerations that lead us to the main result in Eq.~\eqref{eq_EOS} and its consequences in Eq.~\eqref{eq_dPf}, we can repeat the analysis for an ideal gas of massless Bosons (with chemical potential $\mu_B = 0$), whose Grand-potential is
\bea
\Omega_0^{B} &=& -T_0\ln Z_{0}^{B} = -\frac{\mathcal{V} T_0^4}{6 \pi^2}\int_0^{\infty}dx\,\frac{x^3}{e^x -1}\nn\\ 
&=& -\nu_B\mathcal{V}\frac{\pi^2 T_0^4}{90},
\eea
where $\nu_B$ represents the total number of discrete degrees of freedom. The ideal gas pressure for Bosons in equilibrium at temperature $T_0$ is thus
\bea
\mathcal{P}_\text{ig}^{B} = \nu_B\frac{\pi^2 T_0^4}{90},
\label{eq_Pigb}
\eea
while the corresponding excess pressure due to non-equilibrium thermal fluctuations will be, after Eq.~\eqref{eq_dPf}
\bea
\delta \mathcal{P}^{B} &=& \mathcal{P} - \mathcal{P}_\text{ig}^{B} = \frac{\Delta_{\beta}}{2\beta_0\mathcal{V}}\frac{\partial^2}{\partial\beta_0^2}\ln Z_{0}^B\nn\\
&=& \nu_B\frac{\pi^2}{15}\Delta_{\beta} \beta_0^{-6} = \nu_B \frac{\pi^2}{15}\Delta\,T_0^2 > 0,
\eea
a positive quantity as well, in agreement with the general proof presented in Eq.~\eqref{eq_ineq}.

When considering finite values of the chemical potential, the corresponding excess pressure for the Fermion gas calculated from Eq.~\eqref{eq:dP} is illustrated in Fig.~\ref{fig:dP2}. Clearly, the excess pressure increases monotonically with $\mu$, and hence so does the total pressure of the system. This effect can be attributed to the higher density of the medium at finite chemical potential, which intensifies the impact of thermal fluctuations. This phenomenon finds an explanation through the effective-interaction interpretation in the replica formalism: a non-vanishing chemical potential implies a higher density and hence a higher number of interacting particles capable of communicating the existence of thermal fluctuations across various subregions of the system. Consequently, since $\delta\mathcal{P}$ was proven to be positive by a general argument, the total pressure must increase with respect to the ideal gas in thermal equilibrium. This correction, being proportional to the standard deviation in the temperature distribution $\widetilde{\Delta}$ across the ensemble of subsystems, is thus expected to be small but possibly of physical importance in certain scenarios, such as the deconfinement transition to be discussed in the next subsection.

%---------------------------------------------------
\subsection*{Implications for the deconfinement transition}
%---------------------------------------------------
It is interesting to explore the consequences of these results in the context of the deconfinement transition, within the simple bag model considerations. Assuming that the Hadronic phase is mainly constituted by pions (with $\mu = 0$ and $\nu_B = 3$ for charged states $0,\pm$), applying Eq.~\eqref{eq_Pigb} we have that its pressure, including the excess pressure effect due to temperature fluctuations, would be
\be
\mathcal{P}_{\rm{Had}} = 3\frac{\pi^2 T_0^4}{90} + \delta\mathcal{P}^{\rm{Had}}.
\label{eq_Phad}
\ee

On the other hand, for the plasma phase we have $\nu_F = 2\cdot 3\cdot 2 = 12$ for quarks, and $\nu_B = 2\cdot(3^2 - 1) = 16$ for gluons, such that
\bea
\mathcal{P}_{\rm{Plasma}} &=& \left( \nu_B + \frac{7}{4}\nu_F \right)\frac{\pi^2 T_0^4}{90} + \delta \mathcal{P}^{\rm{Plasma}} - B\nn\\
&=& \frac{37\pi^2}{90}T_0^4 + \delta\mathcal{P}^{\rm{Plasma}} - B
\label{eq_Pplasma}
\eea

Here, we included the bag constant $B \sim 200\,$~MeV~\cite{le2000thermal}, and the excess pressure due to temperature fluctuations associated to both quarks and gluons $\delta\mathcal{P}^{\rm{Plasma}} = \delta\mathcal{P}^{G} + \delta\mathcal{P}^{Q} > 0$.

The critical temperature $T_c$ is obtained by imposing the condition of equal pressures at both phases at the phase transition, i.e.
\bea
3\frac{\pi^2 T_c^4}{90} = \frac{37\pi^2}{90}T_c^4 + \delta\mathcal{P}^{\rm{Net}} - B,
\label{eq_Equil}
\eea
where we defined the net excess pressure as
\bea
\delta\mathcal{P}^{\rm{Net}} &=& \delta\mathcal{P}^{\rm{Plasma}} - \delta\mathcal{P}^{\rm{Had}} = \delta\mathcal{P}^{G} - \delta\mathcal{P}^{\rm{Had}} + \delta\mathcal{P}^{Q} \nn\\
&=& 13\frac{\pi^2}{15}\Delta\,T_0^2 + \delta\mathcal{P}^{Q} > 0,
\eea
which is clearly a positive definite quantity.

Finally, solving for $T_c$ in Eq.~\eqref{eq_Equil}, we obtain
\be
T_c = T_c^{0}\left( 1 - \frac{\delta\mathcal{P}^{\rm{Net}}}{\left(T_c^{0}\right)^4} \right)^{1/4} \le T_c^{0},
\ee
with $T_c^{0} = \left( 45 B /17\pi^2 \right)^{1/4} \sim 144\,$~MeV~\cite{le2000thermal} the critical temperature for a homogenous thermalized system. Therefore, we conclude that non-equilibrium temperature fluctuations will in principle decrease the critical temperature for the deconfinement transition.
%----------------------------------------------------------------
\section{Summary and Conclusions}
As an approximation to the non-equilibrium conditions arising in several relativistic quantum systems, such as heavy-ion collisions, we have considered an ensemble of subsystems at different temperatures $T = T_0 + \delta T$, with average $T_0$ and standard deviation $\overline{\delta T^2} = \Delta$. These statistical properties imply that the inverse temperature $\beta = \beta_0 + \delta\beta$ can be modeled by a Gaussian distributed fluctuation $\delta\beta$, with zero mean and standard deviation $\overline{\delta\beta^2} = \Delta_{\beta} = \beta_0^4\Delta$. By first considering a non-interacting system of QED fermions, we applied the replica trick to obtain the statistical average of the Grand Potential as a series expansion at all orders in the parameter $\Delta$. Furthermore, from this exact expression, we obtained the excess pressure with respect to the ideal Fermi gas due to the thermal fluctuations. The same analysis can be carried out for an ideal gas of Bosons, for which we also obtained explicit results for the corresponding excess pressure.
In agreement with our previous works~\cite{PhysRevD.107.096014,PhysRevD.108.116013}, the statistical average over fluctuating parameters (in this case the temperature) within the replica formalism, can be interpreted as an effective particle-particle interaction introduced to the free Lagrangian~\cite{PhysRevD.107.096014,PhysRevD.108.116013}. The strength of these interactions is here proportional to the parameter $\Delta$, that represents the auto-correlation in the temperature fluctuations. 
This result may be of significant impact in the interpretation of non-equilibrium effects and thermal fluctuations in several high-energy systems, particularly on the deconfinement transition between hadronic matter and the quark gluon plasma, for which we provided a simple but straightforward analysis based on the bag model, showing that the critical temperature decreases due to such non-equilibrium fluctuations.

%----------------------------------------------------------------
\acknowledgements{E.M. acknowledges financial support from Fondecyt 1230440. J.D.C.-Y. acknowledges financial support from Fondecyt 3220087. M. L. acknowledges financial support from ANID/CONICYT FONDECYT Regular (Chile) under Grants No. 1200483, No. 1190192, and No. 1220035. }

%\bibliography
%

\appendix

\section{Matsubara sums}
\label{app_Matsubara}
Here we present the detail of the evaluation of the Matsubara sum of the logarithmic functions in the main text.
For simplicity, let us define $\xi_p = \mu \pm E_\mathbf{p}$, and hence consider the generic sum for $\omega_k = (2 k + 1)\pi/\beta_0$,
\bea
S = \sum_{k\in\mathbb{Z}}\ln\left( \ii\omega_k - \xi_p \right).
\eea

\begin{figure}[h!]
\centering
     \includegraphics[scale=1]{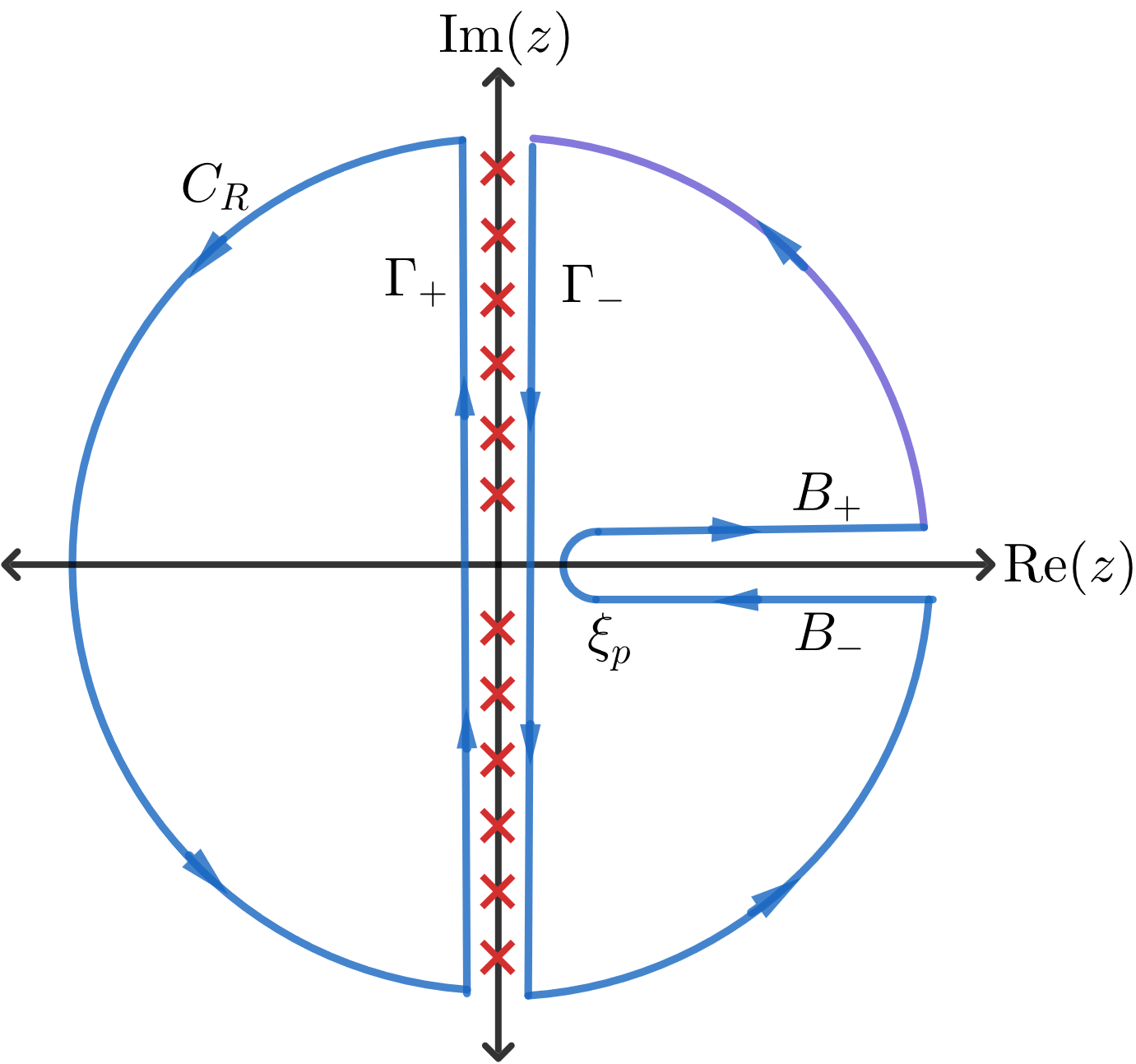}
     \caption{Integration Contour in Eq.~\eqref{eq_int_contour}}
     \label{fig:contour}
 \end{figure}

To evaluate the sum, we shall construct an integration path on the complex contour, by sugin the meromorphic function
\bea
g(z) = \frac{\beta_0}{e^{\beta_0 z} + 1},
\eea
that possesses infinitely many single poles along the imaginary axis at the Matsubara frequencies $z_k =\ii\omega_k$, with residue 1:
\bea
\left.{\rm{Res}}\,g(z)\right|_{z=\ii\omega_k} &=& \lim_{z\rightarrow \ii\omega_k}(z - \ii\omega_k)g(z)\nn\\
&=& \beta_0\lim_{z\rightarrow \ii\omega_k}\frac{z - \ii\omega_k}{1 + e^{\ii\beta_0\omega_k}e^{\beta_0(z - \ii\omega_k)}}\nn\\
&=& \beta_0\lim_{z\rightarrow \ii\omega_k}\frac{z - \ii\omega_k}{1  -e^{\beta_0(z - \ii\omega_k)}}\nn\\
&=& 1
\label{eq_res_g}
\eea

Therefore, we consider the complex integral over the contour illustrated in Fig.~\ref{fig:contour}
\bea
\frac{1}{2\pi \ii}\oint dz\,g(z) \ln(z - \xi_p)&=& \int_{C_R}\frac{dz}{2\pi\ii } g(z) \ln(z - \xi_p)\nn\\ 
&+& 
\int_{B_+ \cup B_-}\frac{dz}{2\pi\ii } g(z) \ln(z - \xi_p)\nn\\
&+& \int_{\Gamma_+ \cup \Gamma_-}\frac{dz}{2\pi\ii } g(z) \ln(z - \xi_p)\nn\\
&=& 0
\label{eq_int_contour}
\eea
where we have surrounded the branch cut of the logarithm starting at $z = \xi_p$, and no poles are enclosed inside the contour. Now, we calculate separately each component.
Clearly, after the exponential contribution in the denominator of $g(z)$, we have
\be
\lim_{R\rightarrow\infty}\int_{C_R}\frac{dz}{2\pi\ii } g(z) \ln(z - \xi_p) = 0.
\ee

The integral that surrounds the Matsubara poles in the imaginary axis is calculated using the residue theorem
\bea
&&\lim_{R\rightarrow\infty}\int_{\Gamma_+ \cup \Gamma_-}\frac{dz}{2\pi\ii } g(z) \ln(z - \xi_p)\nn\\ 
&&= -\sum_{k\in\mathbb{Z}}\lim_{z\rightarrow \ii\omega_k} (z - \ii\omega_k)g(z) \ln(z - \xi_p)\nn\\
&&= -\sum_{k\in\mathbb{Z}}  \ln(\ii\omega_k - \xi_p) = - S,
\label{eq_sum}
\eea
where we applied Eq.~\eqref{eq_res_g} for the residues of $g(z)$.
Substituting Eq.~\eqref{eq_sum} into Eq.~\eqref{eq_int_contour}, we have that the Matsubara sum is given by the integral around the branch cut
\begin{widetext}
\bea
S &=& \sum_{k\in\mathbb{Z}}  \ln(\ii\omega_k - \xi_p) = \int_{B_+ \cup B_-}\frac{dz}{2\pi\ii } g(z) \ln(z - \xi_p)\nn\\
&=& -\frac{1}{2\pi \ii}\int_{\xi_p}^{\infty}dx\, g(x) \left[ \ln(x - \xi_p + \ii\epsilon^{+}) - \ln(x - \xi_p - \ii\epsilon^{+}) \right]\nn\\
&=& -\frac{1}{2\pi \ii}\int_{\xi_p}^{\infty}dx\, \frac{\beta_0 e^{-\beta_0 x}}{e^{-\beta_0 x}+1} \left[ \ln(x - \xi_p + \ii\epsilon^{+}) - \ln(x - \xi_p - \ii\epsilon^{+}) \right]\nn\\
&=& \frac{1}{2\pi \ii}\int_{\xi_p}^{\infty}dx\, \frac{\partial}{\partial x}\ln\left(e^{-\beta_0 x}+1\right) \left[ \ln(x - \xi_p + \ii\epsilon^{+}) - \ln(x - \xi_p - \ii\epsilon^{+}) \right]\nn\\
&=& -\frac{1}{2\pi \ii}\int_{\xi_p}^{\infty}dx\, \ln\left(e^{-\beta_0 x}+1\right) \left[ \frac{1}{x - \xi_p + \ii\epsilon^{+}} - \frac{1}{x - \xi_p - \ii\epsilon^{+}} \right]\nn\\
&=& \int_{\xi_p}^{\infty}dx\, \ln\left(e^{-\beta_0 x}+1\right)\delta(x - \xi_p)\nn\\
&=& \ln\left(e^{-\beta_0 \xi_p}+1\right)
\eea
where in the fourth line we integrated by parts, and in the third step we used the identity
\be
\lim_{\epsilon\rightarrow 0^{+}}\frac{1}{A \pm \ii\epsilon} = PV(1/A) \mp \ii\pi \delta(A)
\ee
\end{widetext}

\section{Thermodynamic relations for the fluctuations}
\label{app_thermo}
As clearly stated in the main text, the first-order in $\Delta$ contribution to the excess pressure is proportional to the second derivative of the Grand Partition function of the reference system, constituted by the relativistic Fermi gas. Using the relation $\beta_0 = T_0^{-1}$, and the definition $\ln Z_0 = - \Omega_0/T_0$, we have
\bea
\frac{\partial^2}{\partial\beta_0^2}\ln Z_0 &=& -T_0\frac{\partial}{\partial T_0}\left( T_0\left.\frac{\partial \Omega_0}{\partial T_0}\right|_{\mu,\mathcal{V}} - \Omega_0 \right)_{\mu,\mathcal{V}}\nn\\
&=& -T_0^3 \left.\frac{\partial^2\Omega_0}{\partial T_0^2}\right|_{\mu,\mathcal{V}}.
\label{eq_der2lnZ}
\eea

On the other hand, from the differential form of the Grand Potential,
\bea
d\Omega_0 = -\mathcal{P}d\mathcal{V} - S dT_0 - Nd\mu,
\eea
we have the relations
\bea
S &=& -\left.\frac{\partial\Omega_0}{\partial T_0}\right|_{\mu,\mathcal{V}},\nn\\
\left.\frac{\partial S}{\partial\mu}\right|_{T_0,\mathcal{V}} &=& \left.\frac{\partial N}{\partial T_0}\right|_{\mu,\mathcal{V}}.
\label{eq_thermrel}
\eea

Inserting the first expression into Eq.~\eqref{eq_der2lnZ}, we obtain
\bea
\frac{\partial^2}{\partial\beta_0^2}\ln Z_0 = T_0^3 \left.\frac{\partial S}{\partial T_0}\right|_{\mu,\mathcal{V}}.
\label{eq_derS}
\eea

We remark that the entropy derivative in Eq.~\eqref{eq_derS} is related to the specific heat at constant volume $C_v$, and hence a positive definite quantity, as we show as follows. By definition, we have (using the Jacobian notation)
\bea
C_{v} &=& T_0\left.\frac{\partial S}{\partial T_0}\right|_{\mathcal{V},N} = T_0\frac{\partial (S,N)}{\partial (T_0,N)} = T_0\frac{\frac{\partial (S,N)}{\partial(T_0,\mu)}}{\frac{\partial (T_0,N)}{\partial(T_0,\mu)}}\nn\\
&=& \frac{T_0}{\left.\frac{\partial N}{\partial \mu}\right|_{T_0,\mathcal{V}}} \left|\begin{array}{cc} \left.\frac{\partial S}{\partial T_0}\right|_{\mu,\mathcal{V}} & \left.\frac{\partial S}{\partial \mu}\right|_{T_0,\mathcal{V}}\\ \left.\frac{\partial N}{\partial T_0}\right|_{\mu,\mathcal{V}} & \left.\frac{\partial N}{\partial \mu}\right|_{T_0,\mathcal{V}} \end{array}\right|
\eea

Evaluating the determinant, we obtain after some elementary algebra
\bea
\frac{C_v}{T_0} &=& \left.\frac{\partial S}{\partial T_0}\right|_{\mu,\mathcal{V}} - \frac{\left.\frac{\partial S}{\partial \mu}\right|_{T_0,\mathcal{V}}\cdot \left.\frac{\partial N}{\partial T_0}\right|_{\mu,\mathcal{V}}}{\left.\frac{\partial N}{\partial \mu}\right|_{T_0,\mathcal{V}}}\nn\\
&=& \left.\frac{\partial S}{\partial T_0}\right|_{\mu,\mathcal{V}}
- \frac{\left(\left.\frac{\partial N}{\partial T_0}\right|_{\mu,\mathcal{V}}\right)^2}{\left.\frac{\partial N}{\partial \mu}\right|_{T_0,\mathcal{V}}},
\label{eq_cv}
\eea
where in the second line we substituted the second relation in Eq.~\eqref{eq_thermrel}. 
From Eq.~\eqref{eq_cv}, we obtain
\bea
\left.\frac{\partial S}{\partial T_0}\right|_{\mu,\mathcal{V}} = 
\frac{C_v}{T_0}
+ \frac{\left(\left.\frac{\partial N}{\partial T_0}\right|_{\mu,\mathcal{V}}\right)^2}{\left.\frac{\partial N}{\partial \mu}\right|_{T_0,\mathcal{V}}}
\eea
Therefore, substituting this result into Eq.~\eqref{eq_derS}, we obtain
\bea
\frac{\partial^2}{\partial\beta_0^2}\ln Z_0 &=& T_0^2 \left( C_v + T_0\frac{\left(\left.\frac{\partial N}{\partial T_0}\right|_{\mu,\mathcal{V}}\right)^2}{\left.\frac{\partial N}{\partial \mu}\right|_{T_0,\mathcal{V}}} \right)
\label{eq_derln3}
\eea

Finally, applying the statistical-mechanical definition fo the average particle number $N = \langle \hat{N} \rangle$ in the Grand Canonical Ensemble, we have
\bea
\left.\frac{\partial N}{\partial \mu}\right|_{T_0,\mathcal{V}} = \frac{1}{T_0}\left(  \langle\hat{N}^2\rangle - \langle\hat{N}\rangle^2\right) = \frac{\langle (\delta\hat{N})^2\rangle}{T_0}\ge 0,
\eea
which combined with Eq.~\eqref{eq_derln3} leads us to prove the inequality
\bea
\frac{\partial^2}{\partial\beta_0^2}\ln Z_0 &=& T_0^3 \left.\frac{\partial S}{\partial T_0}\right|_{\mu,\mathcal{V}} \\
&=&  T_0^2 \left( C_v + T_0^2\frac{\left(\left.\frac{\partial N}{\partial T_0}\right|_{\mu,\mathcal{V}}\right)^2}{\langle (\delta\hat{N})^2\rangle} \right)
\ge 0,\nn
\eea
as stated in the main text.

\begin{thebibliography}{29}%
\makeatletter
\providecommand \@ifxundefined [1]{%
 \@ifx{#1\undefined}
}%
\providecommand \@ifnum [1]{%
 \ifnum #1\expandafter \@firstoftwo
 \else \expandafter \@secondoftwo
 \fi
}%
\providecommand \@ifx [1]{%
 \ifx #1\expandafter \@firstoftwo
 \else \expandafter \@secondoftwo
 \fi
}%
\providecommand \natexlab [1]{#1}%
\providecommand \enquote  [1]{``#1''}%
\providecommand \bibnamefont  [1]{#1}%
\providecommand \bibfnamefont [1]{#1}%
\providecommand \citenamefont [1]{#1}%
\providecommand \href@noop [0]{\@secondoftwo}%
\providecommand \href [0]{\begingroup \@sanitize@url \@href}%
\providecommand \@href[1]{\@@startlink{#1}\@@href}%
\providecommand \@@href[1]{\endgroup#1\@@endlink}%
\providecommand \@sanitize@url [0]{\catcode `\\12\catcode `\$12\catcode `\&12\catcode `\#12\catcode `\^12\catcode `\_12\catcode `\%12\relax}%
\providecommand \@@startlink[1]{}%
\providecommand \@@endlink[0]{}%
\providecommand \url  [0]{\begingroup\@sanitize@url \@url }%
\providecommand \@url [1]{\endgroup\@href {#1}{\urlprefix }}%
\providecommand \urlprefix  [0]{URL }%
\providecommand \Eprint [0]{\href }%
\providecommand \doibase [0]{http://dx.doi.org/}%
\providecommand \selectlanguage [0]{\@gobble}%
\providecommand \bibinfo  [0]{\@secondoftwo}%
\providecommand \bibfield  [0]{\@secondoftwo}%
\providecommand \translation [1]{[#1]}%
\providecommand \BibitemOpen [0]{}%
\providecommand \bibitemStop [0]{}%
\providecommand \bibitemNoStop [0]{.\EOS\space}%
\providecommand \EOS [0]{\spacefactor3000\relax}%
\providecommand \BibitemShut  [1]{\csname bibitem#1\endcsname}%
\let\auto@bib@innerbib\@empty
%</preamble>
\bibitem [{\citenamefont {Altland}\ and\ \citenamefont {Simons}(2010)}]{altland2010condensed}%
  \BibitemOpen
  \bibfield  {author} {\bibinfo {author} {\bibfnamefont {Alexander}\ \bibnamefont {Altland}}\ and\ \bibinfo {author} {\bibfnamefont {Ben~D}\ \bibnamefont {Simons}},\ }\href@noop {} {\emph {\bibinfo {title} {Condensed matter field theory}}}\ (\bibinfo  {publisher} {Cambridge university press},\ \bibinfo {year} {2010})\BibitemShut {NoStop}%
\bibitem [{\citenamefont {Le~Bellac}(2000)}]{le2000thermal}%
  \BibitemOpen
  \bibfield  {author} {\bibinfo {author} {\bibfnamefont {Michel}\ \bibnamefont {Le~Bellac}},\ }\href@noop {} {\emph {\bibinfo {title} {Thermal field theory}}}\ (\bibinfo  {publisher} {Cambridge university press},\ \bibinfo {year} {2000})\BibitemShut {NoStop}%
\bibitem [{\citenamefont {Kapusta}\ and\ \citenamefont {Gale}(2007)}]{kapusta2007finite}%
  \BibitemOpen
  \bibfield  {author} {\bibinfo {author} {\bibfnamefont {Joseph~I}\ \bibnamefont {Kapusta}}\ and\ \bibinfo {author} {\bibfnamefont {Charles}\ \bibnamefont {Gale}},\ }\href@noop {} {\emph {\bibinfo {title} {Finite-temperature field theory: Principles and applications}}}\ (\bibinfo  {publisher} {Cambridge university press},\ \bibinfo {year} {2007})\BibitemShut {NoStop}%
\bibitem [{\citenamefont {Arsene}\ \emph {et~al.}(2005)\citenamefont {Arsene} \emph {et~al.}}]{ARSENE20051}%
  \BibitemOpen
  \bibfield  {author} {\bibinfo {author} {\bibfnamefont {I.}~\bibnamefont {Arsene}} \emph {et~al.} (\bibinfo {collaboration} {BRAHMS Collaboration}),\ }\bibfield  {title} {\enquote {\bibinfo {title} {{Quark–gluon plasma and color glass condensate at RHIC? The perspective from the BRAHMS experiment}},}\ }\href {\doibase https://doi.org/10.1016/j.nuclphysa.2005.02.130} {\bibfield  {journal} {\bibinfo  {journal} {Nuc. Phys. A}\ }\textbf {\bibinfo {volume} {757}},\ \bibinfo {pages} {1--27} (\bibinfo {year} {2005})}\BibitemShut {NoStop}%
\bibitem [{\citenamefont {Adcox}\ \emph {et~al.}(2005)\citenamefont {Adcox} \emph {et~al.}}]{ADCOX2005184}%
  \BibitemOpen
  \bibfield  {author} {\bibinfo {author} {\bibfnamefont {K.}~\bibnamefont {Adcox}} \emph {et~al.} (\bibinfo {collaboration} {PHENIX Collaboration}),\ }\bibfield  {title} {\enquote {\bibinfo {title} {{Formation of dense partonic matter in relativistic nucleus–nucleus collisions at RHIC: Experimental evaluation by the PHENIX Collaboration}},}\ }\href {\doibase https://doi.org/10.1016/j.nuclphysa.2005.03.086} {\bibfield  {journal} {\bibinfo  {journal} {Nuc. Phys. A}\ }\textbf {\bibinfo {volume} {757}},\ \bibinfo {pages} {184--283} (\bibinfo {year} {2005})}\BibitemShut {NoStop}%
\bibitem [{\citenamefont {Becattini}(2014)}]{Becattini_2014}%
  \BibitemOpen
  \bibfield  {author} {\bibinfo {author} {\bibfnamefont {Francesco}\ \bibnamefont {Becattini}},\ }\bibfield  {title} {\enquote {\bibinfo {title} {{The Quark Gluon Plasma and relativistic heavy ion collisions in the LHC era}},}\ }\href {\doibase 10.1088/1742-6596/527/1/012012} {\bibfield  {journal} {\bibinfo  {journal} {Journal of Physics: Conference Series}\ }\textbf {\bibinfo {volume} {527}},\ \bibinfo {pages} {012012} (\bibinfo {year} {2014})}\BibitemShut {NoStop}%
\bibitem [{\citenamefont {Florkowski}\ and\ \citenamefont {Ryblewski}(2011)}]{PhysRevC.83.034907}%
  \BibitemOpen
  \bibfield  {author} {\bibinfo {author} {\bibfnamefont {Wojciech}\ \bibnamefont {Florkowski}}\ and\ \bibinfo {author} {\bibfnamefont {Radoslaw}\ \bibnamefont {Ryblewski}},\ }\bibfield  {title} {\enquote {\bibinfo {title} {Highly anisotropic and strongly dissipative hydrodynamics for early stages of relativistic heavy-ion collisions},}\ }\href {\doibase 10.1103/PhysRevC.83.034907} {\bibfield  {journal} {\bibinfo  {journal} {Phys. Rev. C}\ }\textbf {\bibinfo {volume} {83}},\ \bibinfo {pages} {034907} (\bibinfo {year} {2011})}\BibitemShut {NoStop}%
\bibitem [{\citenamefont {Blaizot}\ and\ \citenamefont {Mueller}(1987)}]{BLAIZOT1987847}%
  \BibitemOpen
  \bibfield  {author} {\bibinfo {author} {\bibfnamefont {J.P.}\ \bibnamefont {Blaizot}}\ and\ \bibinfo {author} {\bibfnamefont {A.H.}\ \bibnamefont {Mueller}},\ }\bibfield  {title} {\enquote {\bibinfo {title} {The early stage of ultra-relativistic heavy ion collisions},}\ }\href {\doibase https://doi.org/10.1016/0550-3213(87)90408-1} {\bibfield  {journal} {\bibinfo  {journal} {Nucl. Phys. B.}\ }\textbf {\bibinfo {volume} {289}},\ \bibinfo {pages} {847--860} (\bibinfo {year} {1987})}\BibitemShut {NoStop}%
\bibitem [{\citenamefont {Zschocke}\ \emph {et~al.}(2011)\citenamefont {Zschocke}, \citenamefont {Horv\'at}, \citenamefont {Mishustin},\ and\ \citenamefont {Csernai}}]{PhysRevC.83.044903}%
  \BibitemOpen
  \bibfield  {author} {\bibinfo {author} {\bibfnamefont {Sven}\ \bibnamefont {Zschocke}}, \bibinfo {author} {\bibfnamefont {Szabolcs}\ \bibnamefont {Horv\'at}}, \bibinfo {author} {\bibfnamefont {Igor~N.}\ \bibnamefont {Mishustin}}, \ and\ \bibinfo {author} {\bibfnamefont {L\'aszl\'o~P.}\ \bibnamefont {Csernai}},\ }\bibfield  {title} {\enquote {\bibinfo {title} {Nonequilibrium hadronization and constituent quark number scaling},}\ }\href {\doibase 10.1103/PhysRevC.83.044903} {\bibfield  {journal} {\bibinfo  {journal} {Phys. Rev. C}\ }\textbf {\bibinfo {volume} {83}},\ \bibinfo {pages} {044903} (\bibinfo {year} {2011})}\BibitemShut {NoStop}%
\bibitem [{\citenamefont {Rafelski}\ and\ \citenamefont {Letessier}(2002)}]{RafelskyHadronization}%
  \BibitemOpen
  \bibfield  {author} {\bibinfo {author} {\bibfnamefont {Johann}\ \bibnamefont {Rafelski}}\ and\ \bibinfo {author} {\bibfnamefont {Jean}\ \bibnamefont {Letessier}},\ }\bibfield  {title} {\enquote {\bibinfo {title} {{{Non‐equilibrium Hadrochemistry in QGP Hadronization}}},}\ }\href {\doibase 10.1063/1.1513695} {\bibfield  {journal} {\bibinfo  {journal} {AIP Conference Proceedings}\ }\textbf {\bibinfo {volume} {631}},\ \bibinfo {pages} {460--489} (\bibinfo {year} {2002})}\BibitemShut {NoStop}%
\bibitem [{\citenamefont {Keldysh}()}]{doi:10.1142/9789811279461_0007}%
  \BibitemOpen
  \bibfield  {author} {\bibinfo {author} {\bibfnamefont {L.~V.}\ \bibnamefont {Keldysh}},\ }\enquote {\bibinfo {title} {Diagram technique for nonequilibrium processes},}\ in\ \href {\doibase 10.1142/9789811279461_0007} {\emph {\bibinfo {booktitle} {Selected Papers of Leonid V Keldysh}}},\ pp.\ \bibinfo {pages} {47--55}\BibitemShut {NoStop}%
\bibitem [{\citenamefont {Álamo}\ and\ \citenamefont {Muñoz}(2018)}]{e20050366}%
  \BibitemOpen
  \bibfield  {author} {\bibinfo {author} {\bibfnamefont {Manuel}\ \bibnamefont {Álamo}}\ and\ \bibinfo {author} {\bibfnamefont {Enrique}\ \bibnamefont {Muñoz}},\ }\bibfield  {title} {\enquote {\bibinfo {title} {Thermoelectric efficiency of a topological nano-junction},}\ }\href {\doibase 10.3390/e20050366} {\bibfield  {journal} {\bibinfo  {journal} {Entropy}\ }\textbf {\bibinfo {volume} {20}} (\bibinfo {year} {2018}),\ 10.3390/e20050366}\BibitemShut {NoStop}%
\bibitem [{\citenamefont {Mu\~noz}\ \emph {et~al.}(2013)\citenamefont {Mu\~noz}, \citenamefont {Bolech},\ and\ \citenamefont {Kirchner}}]{PhysRevLett.110.016601}%
  \BibitemOpen
  \bibfield  {author} {\bibinfo {author} {\bibfnamefont {Enrique}\ \bibnamefont {Mu\~noz}}, \bibinfo {author} {\bibfnamefont {C.~J.}\ \bibnamefont {Bolech}}, \ and\ \bibinfo {author} {\bibfnamefont {Stefan}\ \bibnamefont {Kirchner}},\ }\bibfield  {title} {\enquote {\bibinfo {title} {{Universal Out-of-Equilibrium Transport in Kondo-Correlated Quantum Dots: Renormalized Dual Fermions on the Keldysh Contour}},}\ }\href {\doibase 10.1103/PhysRevLett.110.016601} {\bibfield  {journal} {\bibinfo  {journal} {Phys. Rev. Lett.}\ }\textbf {\bibinfo {volume} {110}},\ \bibinfo {pages} {016601} (\bibinfo {year} {2013})}\BibitemShut {NoStop}%
\bibitem [{\citenamefont {Muñoz}\ \emph {et~al.}(2017)\citenamefont {Muñoz}, \citenamefont {Zamani}, \citenamefont {Merker}, \citenamefont {Costi},\ and\ \citenamefont {Kirchner}}]{Muñoz_2017}%
  \BibitemOpen
  \bibfield  {author} {\bibinfo {author} {\bibfnamefont {Enrique}\ \bibnamefont {Muñoz}}, \bibinfo {author} {\bibfnamefont {Farzaneh}\ \bibnamefont {Zamani}}, \bibinfo {author} {\bibfnamefont {Lukas}\ \bibnamefont {Merker}}, \bibinfo {author} {\bibfnamefont {Theo}\ \bibnamefont {Costi}}, \ and\ \bibinfo {author} {\bibfnamefont {Stefan}\ \bibnamefont {Kirchner}},\ }\bibfield  {title} {\enquote {\bibinfo {title} {The renormalized superperturbation theory (rspt) approach to the anderson model in and out of equilibrium},}\ }\href {\doibase 10.1088/1742-6596/807/9/092001} {\bibfield  {journal} {\bibinfo  {journal} {Journal of Physics: Conference Series}\ }\textbf {\bibinfo {volume} {807}},\ \bibinfo {pages} {092001} (\bibinfo {year} {2017})}\BibitemShut {NoStop}%
\bibitem [{\citenamefont {Falomir}\ \emph {et~al.}(2018)\citenamefont {Falomir}, \citenamefont {Loewe}, \citenamefont {Mu\~noz},\ and\ \citenamefont {Raya}}]{PhysRevB.98.195430}%
  \BibitemOpen
  \bibfield  {author} {\bibinfo {author} {\bibfnamefont {Horacio}\ \bibnamefont {Falomir}}, \bibinfo {author} {\bibfnamefont {Marcelo}\ \bibnamefont {Loewe}}, \bibinfo {author} {\bibfnamefont {Enrique}\ \bibnamefont {Mu\~noz}}, \ and\ \bibinfo {author} {\bibfnamefont {Alfredo}\ \bibnamefont {Raya}},\ }\bibfield  {title} {\enquote {\bibinfo {title} {Optical conductivity and transparency in an effective model for graphene},}\ }\href {\doibase 10.1103/PhysRevB.98.195430} {\bibfield  {journal} {\bibinfo  {journal} {Phys. Rev. B}\ }\textbf {\bibinfo {volume} {98}},\ \bibinfo {pages} {195430} (\bibinfo {year} {2018})}\BibitemShut {NoStop}%
\bibitem [{\citenamefont {Kirchner}\ \emph {et~al.}(2013)\citenamefont {Kirchner}, \citenamefont {Zamani},\ and\ \citenamefont {Mu{\~{n}}oz}}]{MuñozBook}%
  \BibitemOpen
  \bibfield  {author} {\bibinfo {author} {\bibfnamefont {Stefan}\ \bibnamefont {Kirchner}}, \bibinfo {author} {\bibfnamefont {Farzaneh}\ \bibnamefont {Zamani}}, \ and\ \bibinfo {author} {\bibfnamefont {Enrique}\ \bibnamefont {Mu{\~{n}}oz}},\ }\bibfield  {title} {\enquote {\bibinfo {title} {{Nonlinear Thermoelectric Response of Quantum Dots: Renormalized Dual Fermions Out of Equilibrium}},}\ }in\ \href {\doibase https://doi.org/10.1007/978-94-007-4984-9_10} {\emph {\bibinfo {booktitle} {\small New Materials for Thermoelectric \\ Applications: Theory and Experiment}}}\ (\bibinfo  {publisher} {Springer Netherlands},\ \bibinfo {address} {Dordrecht},\ \bibinfo {year} {2013})\ pp.\ \bibinfo {pages} {129--168}\BibitemShut {NoStop}%
\bibitem [{\citenamefont {Aoki}\ \emph {et~al.}(2014)\citenamefont {Aoki}, \citenamefont {Tsuji}, \citenamefont {Eckstein}, \citenamefont {Kollar}, \citenamefont {Oka},\ and\ \citenamefont {Werner}}]{RevModPhys.86.779}%
  \BibitemOpen
  \bibfield  {author} {\bibinfo {author} {\bibfnamefont {Hideo}\ \bibnamefont {Aoki}}, \bibinfo {author} {\bibfnamefont {Naoto}\ \bibnamefont {Tsuji}}, \bibinfo {author} {\bibfnamefont {Martin}\ \bibnamefont {Eckstein}}, \bibinfo {author} {\bibfnamefont {Marcus}\ \bibnamefont {Kollar}}, \bibinfo {author} {\bibfnamefont {Takashi}\ \bibnamefont {Oka}}, \ and\ \bibinfo {author} {\bibfnamefont {Philipp}\ \bibnamefont {Werner}},\ }\bibfield  {title} {\enquote {\bibinfo {title} {Nonequilibrium dynamical mean-field theory and its applications},}\ }\href {\doibase 10.1103/RevModPhys.86.779} {\bibfield  {journal} {\bibinfo  {journal} {Rev. Mod. Phys.}\ }\textbf {\bibinfo {volume} {86}},\ \bibinfo {pages} {779--837} (\bibinfo {year} {2014})}\BibitemShut {NoStop}%
\bibitem [{\citenamefont {Sieberer}\ \emph {et~al.}(2016)\citenamefont {Sieberer}, \citenamefont {Buchhold},\ and\ \citenamefont {Diehl}}]{Sieberer_2016}%
  \BibitemOpen
  \bibfield  {author} {\bibinfo {author} {\bibfnamefont {L~M}\ \bibnamefont {Sieberer}}, \bibinfo {author} {\bibfnamefont {M}~\bibnamefont {Buchhold}}, \ and\ \bibinfo {author} {\bibfnamefont {S}~\bibnamefont {Diehl}},\ }\bibfield  {title} {\enquote {\bibinfo {title} {Keldysh field theory for driven open quantum systems},}\ }\href {\doibase 10.1088/0034-4885/79/9/096001} {\bibfield  {journal} {\bibinfo  {journal} {Rep. Prog. Phys}\ }\textbf {\bibinfo {volume} {79}},\ \bibinfo {pages} {096001} (\bibinfo {year} {2016})}\BibitemShut {NoStop}%
\bibitem [{\citenamefont {Beck}\ and\ \citenamefont {Cohen}(2003)}]{beck2003superstatistics}%
  \BibitemOpen
  \bibfield  {author} {\bibinfo {author} {\bibfnamefont {Christian}\ \bibnamefont {Beck}}\ and\ \bibinfo {author} {\bibfnamefont {Ezechiel G.~D.}\ \bibnamefont {Cohen}},\ }\bibfield  {title} {\enquote {\bibinfo {title} {Superstatistics},}\ }\href {https://www.sciencedirect.com/science/article/abs/pii/S0378437103000190} {\bibfield  {journal} {\bibinfo  {journal} {Phys. A}\ }\textbf {\bibinfo {volume} {322}},\ \bibinfo {pages} {267--275} (\bibinfo {year} {2003})}\BibitemShut {NoStop}%
\bibitem [{\citenamefont {Sargolzaeipor}\ \emph {et~al.}(2019)\citenamefont {Sargolzaeipor}, \citenamefont {Hassanabadi},\ and\ \citenamefont {Chung}}]{doi:10.1142/S0217732319500238}%
  \BibitemOpen
  \bibfield  {author} {\bibinfo {author} {\bibfnamefont {S.}~\bibnamefont {Sargolzaeipor}}, \bibinfo {author} {\bibfnamefont {H.}~\bibnamefont {Hassanabadi}}, \ and\ \bibinfo {author} {\bibfnamefont {W.~S.}\ \bibnamefont {Chung}},\ }\bibfield  {title} {\enquote {\bibinfo {title} {Superstatistics of two electrons quantum dot},}\ }\href {\doibase 10.1142/S0217732319500238} {\bibfield  {journal} {\bibinfo  {journal} {Modern Physics Letters A}\ }\textbf {\bibinfo {volume} {34}},\ \bibinfo {pages} {1950023} (\bibinfo {year} {2019})}\BibitemShut {NoStop}%
\bibitem [{\citenamefont {Casta{\~n}o-Yepes}\ and\ \citenamefont {Amor-Quiroz}(2020)}]{castano2020super}%
  \BibitemOpen
  \bibfield  {author} {\bibinfo {author} {\bibfnamefont {Jorge~David}\ \bibnamefont {Casta{\~n}o-Yepes}}\ and\ \bibinfo {author} {\bibfnamefont {D.~A.}\ \bibnamefont {Amor-Quiroz}},\ }\bibfield  {title} {\enquote {\bibinfo {title} {Super-statistical description of thermo-magnetic properties of a system of {2D} {GaAs} quantum dots with gaussian confinement and {Rashba} spin--orbit interaction},}\ }\href {https://www.sciencedirect.com/science/article/abs/pii/S0378437119321508} {\bibfield  {journal} {\bibinfo  {journal} {Phys. A}\ }\textbf {\bibinfo {volume} {548}},\ \bibinfo {pages} {123871} (\bibinfo {year} {2020})}\BibitemShut {NoStop}%
\bibitem [{\citenamefont {Ourabah}\ and\ \citenamefont {Tribeche}(2017)}]{PhysRevE.95.042111}%
  \BibitemOpen
  \bibfield  {author} {\bibinfo {author} {\bibfnamefont {Kamel}\ \bibnamefont {Ourabah}}\ and\ \bibinfo {author} {\bibfnamefont {Mouloud}\ \bibnamefont {Tribeche}},\ }\bibfield  {title} {\enquote {\bibinfo {title} {Quantum entanglement and temperature fluctuations},}\ }\href {\doibase 10.1103/PhysRevE.95.042111} {\bibfield  {journal} {\bibinfo  {journal} {Phys. Rev. E}\ }\textbf {\bibinfo {volume} {95}},\ \bibinfo {pages} {042111} (\bibinfo {year} {2017})}\BibitemShut {NoStop}%
\bibitem [{\citenamefont {Casta\~no Yepes}\ and\ \citenamefont {Ramirez-Gutierrez}(2021)}]{PhysRevE.104.024139}%
  \BibitemOpen
  \bibfield  {author} {\bibinfo {author} {\bibfnamefont {Jorge~David}\ \bibnamefont {Casta\~no Yepes}}\ and\ \bibinfo {author} {\bibfnamefont {Cristian~Felipe}\ \bibnamefont {Ramirez-Gutierrez}},\ }\bibfield  {title} {\enquote {\bibinfo {title} {{Superstatistics and quantum entanglement in the isotropic spin-1/2 $XX$ dimer from a nonadditive thermodynamics perspective}},}\ }\href {\doibase 10.1103/PhysRevE.104.024139} {\bibfield  {journal} {\bibinfo  {journal} {Phys. Rev. E}\ }\textbf {\bibinfo {volume} {104}},\ \bibinfo {pages} {024139} (\bibinfo {year} {2021})}\BibitemShut {NoStop}%
\bibitem [{\citenamefont {Castaño-Yepes}(2022)}]{Castaño-Yepes2022}%
  \BibitemOpen
  \bibfield  {author} {\bibinfo {author} {\bibfnamefont {Jorge~David}\ \bibnamefont {Castaño-Yepes}},\ }\bibfield  {title} {\enquote {\bibinfo {title} {Entropy exchange and thermal fluctuations in the jaynes–cummings model},}\ }\href {\doibase 10.1140/epjp/s13360-022-02382-7} {\bibfield  {journal} {\bibinfo  {journal} {Eur. Phys. J. Plus}\ }\textbf {\bibinfo {volume} {137}},\ \bibinfo {pages} {155} (\bibinfo {year} {2022})}\BibitemShut {NoStop}%
\bibitem [{\citenamefont {Ayala}\ \emph {et~al.}(2018)\citenamefont {Ayala}, \citenamefont {Hentschinski}, \citenamefont {Hern\'andez}, \citenamefont {Loewe},\ and\ \citenamefont {Zamora}}]{ayala2018superstatistics}%
  \BibitemOpen
  \bibfield  {author} {\bibinfo {author} {\bibfnamefont {Alejandro}\ \bibnamefont {Ayala}}, \bibinfo {author} {\bibfnamefont {Martin}\ \bibnamefont {Hentschinski}}, \bibinfo {author} {\bibfnamefont {L.~A.}\ \bibnamefont {Hern\'andez}}, \bibinfo {author} {\bibfnamefont {M.}~\bibnamefont {Loewe}}, \ and\ \bibinfo {author} {\bibfnamefont {R.}~\bibnamefont {Zamora}},\ }\bibfield  {title} {\enquote {\bibinfo {title} {{Superstatistics and the effective QCD phase diagram}},}\ }\href {\doibase 10.1103/PhysRevD.98.114002} {\bibfield  {journal} {\bibinfo  {journal} {Phys. Rev. D}\ }\textbf {\bibinfo {volume} {98}},\ \bibinfo {pages} {114002} (\bibinfo {year} {2018})}\BibitemShut {NoStop}%
\bibitem [{\citenamefont {Casta\~no Yepes}\ \emph {et~al.}(2022)\citenamefont {Casta\~no Yepes}, \citenamefont {Mart\'{\i}nez~Paniagua}, \citenamefont {Mu\~noz Vitelly},\ and\ \citenamefont {Ramirez-Gutierrez}}]{PhysRevD.106.116019}%
  \BibitemOpen
  \bibfield  {author} {\bibinfo {author} {\bibfnamefont {Jorge~David}\ \bibnamefont {Casta\~no Yepes}}, \bibinfo {author} {\bibfnamefont {Fernando}\ \bibnamefont {Mart\'{\i}nez~Paniagua}}, \bibinfo {author} {\bibfnamefont {Victor}\ \bibnamefont {Mu\~noz Vitelly}}, \ and\ \bibinfo {author} {\bibfnamefont {Cristian~Felipe}\ \bibnamefont {Ramirez-Gutierrez}},\ }\bibfield  {title} {\enquote {\bibinfo {title} {{Volume effects on the QCD critical end point from thermal fluctuations within the super statistics framework}},}\ }\href {\doibase 10.1103/PhysRevD.106.116019} {\bibfield  {journal} {\bibinfo  {journal} {Phys. Rev. D}\ }\textbf {\bibinfo {volume} {106}},\ \bibinfo {pages} {116019} (\bibinfo {year} {2022})}\BibitemShut {NoStop}%
\bibitem [{\citenamefont {Mezard}\ \emph {et~al.}(1986)\citenamefont {Mezard}, \citenamefont {Parisi},\ and\ \citenamefont {Virasoro}}]{doi:10.1142/0271}%
  \BibitemOpen
  \bibfield  {author} {\bibinfo {author} {\bibfnamefont {M}~\bibnamefont {Mezard}}, \bibinfo {author} {\bibfnamefont {G}~\bibnamefont {Parisi}}, \ and\ \bibinfo {author} {\bibfnamefont {M}~\bibnamefont {Virasoro}},\ }\href {\doibase 10.1142/0271} {\emph {\bibinfo {title} {Spin Glass Theory and Beyond}}}\ (\bibinfo  {publisher} {WORLD SCIENTIFIC},\ \bibinfo {year} {1986})\BibitemShut {NoStop}%
\bibitem [{\citenamefont {Casta\~no Yepes}\ \emph {et~al.}(2023{\natexlab{a}})\citenamefont {Casta\~no Yepes}, \citenamefont {Loewe}, \citenamefont {Mu\~noz}, \citenamefont {Rojas},\ and\ \citenamefont {Zamora}}]{PhysRevD.107.096014}%
  \BibitemOpen
  \bibfield  {author} {\bibinfo {author} {\bibfnamefont {Jorge~David}\ \bibnamefont {Casta\~no Yepes}}, \bibinfo {author} {\bibfnamefont {Marcelo}\ \bibnamefont {Loewe}}, \bibinfo {author} {\bibfnamefont {Enrique}\ \bibnamefont {Mu\~noz}}, \bibinfo {author} {\bibfnamefont {Juan~Crist\'obal}\ \bibnamefont {Rojas}}, \ and\ \bibinfo {author} {\bibfnamefont {Renato}\ \bibnamefont {Zamora}},\ }\bibfield  {title} {\enquote {\bibinfo {title} {{QED fermions in a noisy magnetic field background}},}\ }\href {\doibase 10.1103/PhysRevD.107.096014} {\bibfield  {journal} {\bibinfo  {journal} {Phys. Rev. D}\ }\textbf {\bibinfo {volume} {107}},\ \bibinfo {pages} {096014} (\bibinfo {year} {2023}{\natexlab{a}})}\BibitemShut {NoStop}%
\bibitem [{\citenamefont {Casta\~no Yepes}\ \emph {et~al.}(2023{\natexlab{b}})\citenamefont {Casta\~no Yepes}, \citenamefont {Loewe}, \citenamefont {Mu\~noz},\ and\ \citenamefont {Rojas}}]{PhysRevD.108.116013}%
  \BibitemOpen
  \bibfield  {author} {\bibinfo {author} {\bibfnamefont {Jorge~David}\ \bibnamefont {Casta\~no Yepes}}, \bibinfo {author} {\bibfnamefont {Marcelo}\ \bibnamefont {Loewe}}, \bibinfo {author} {\bibfnamefont {Enrique}\ \bibnamefont {Mu\~noz}}, \ and\ \bibinfo {author} {\bibfnamefont {Juan~Crist\'obal}\ \bibnamefont {Rojas}},\ }\bibfield  {title} {\enquote {\bibinfo {title} {Qed fermions in a noisy magnetic field background: The effective action approach},}\ }\href {\doibase 10.1103/PhysRevD.108.116013} {\bibfield  {journal} {\bibinfo  {journal} {Phys. Rev. D}\ }\textbf {\bibinfo {volume} {108}},\ \bibinfo {pages} {116013} (\bibinfo {year} {2023}{\natexlab{b}})}\BibitemShut {NoStop}%
\end{thebibliography}
\end{document}